\include{amsmath}
\documentclass[9pt,twocolumn,twoside]{osajnl}

\journal{ol} 

\setboolean{shortarticle}{true}

\title{Coherence time of a cold-atom laser below threshold}

\author[1]{Graeme Harvie}
\author[1]{Adam Butcher}
\author[1,*]{Jon Goldwin}

\affil[1]{School of Physics and Astronomy, University of Birmingham, Edgbaston, Birmingham, B15 2TT, United Kingdom}

\affil[*]{Corresponding author: j.m.goldwin@bham.ac.uk}

\begin{abstract}
We experimentally study the coherence time of a below-threshold Raman laser in which the gain medium is a gas of magneto-optically trapped atoms. The second-order optical coherence exhibits photon bunching with a correlation time which is varied by two orders of magnitude by controlling the gain. Results are in good agreement with a simple analytic model which suggests the effect is dominated by gain, rather than dispersion, in this system. Cavity ring-down measurements show the photon lifetime, related to the first-order coherence time, is also increased.
\end{abstract}

\setboolean{displaycopyright}{true}

\begin{document}

\maketitle
A hallmark of quantum optics is the phenomenon of photon bunching observed in the two-time coincidence counts from a thermal light source. The pioneering work of Hanbury Brown and Twiss (HBT) in the context of stellar interferometry \cite{HBT} is now considered a key early milestone in the development of the quantum theory of light in general and of optical coherence in particular \cite{Glauber}. Today photon correlation spectroscopy is in widespread use for characterizing complex media \cite{DLS}, and pseudo-thermal light has been used to perform ghost imaging without the need for entangled photon pairs \cite{ghost}.

Optical amplifiers also can exhibit bunching. Spontaneous emission leads to excess intensity fluctuations, in which case the coherence time reflects the radiative properties of the gain medium, rather than the size or motion of the source. A laser is just an optical amplifier subject to feedback in a resonator, and below the lasing threshold the cavity photon number distribution is identical to that of a thermal state whose effective temperature is dictated by the ratio of gain to losses \cite{QO}. In conventional lasers, the atomic decay rate is much faster than the photon decay rate of the cold (evacuated) cavity, and the atomic dynamics can be adiabatically eliminated. In this regime, known as the good-cavity limit, coherence is enforced through constructive interference of light from multiple round trips. The coherence time therefore scales with the cold cavity lifetime, increasing as the gain approaches threshold \cite{Haken}.

Lasers utilizing cold atomic vapors as gain media can leverage long atomic coherence times to enter the bad-cavity (or good-atom) limit. In this regime, the strong normal dispersion around the narrow gain resonance leads to slow group velocity \cite{vanExter}; above threshold the quantum-limited linewidth is reduced \cite{Kuppens} and there can be a crossover from conventional lasing to steady-state superradiance \cite{Meiser2009,Meiser2010,Mascarenhas,Tieri,Debnath}. A variety of experiments with cold atoms have demonstrated lasing with gain linewidths on the order of, or even much narrower than, the cold cavity bandwidth \cite{Hilico,Guerin,Bohnet,NorciaPRXa,NorciaSci,NorciaPRXb,Megyeri2018a,Gothe,Laske,Schaffer}. Using heterodyne spectroscopy, the first-order coherence time has been measured for bad-cavity lasers using cesium \cite{Hilico}, rubidium \cite{Bohnet}, and strontium atoms \cite{NorciaPRXa,NorciaSci,NorciaPRXb}. The second-order coherence was measured for potassium \cite{Megyeri2018a} and ytterbium \cite{Gothe} atoms as a means to characterize the transition to lasing; both experiments showed bunching below threshold, but a systematic study of the corresponding coherence time remains to be performed. In this letter, we measure the coherence time of photon bunching in a cold-atom ring laser below threshold. Our laser exploits Raman gain which arises during magneto-optical trapping of potassium atoms, in an intermediate regime between the good- and bad-cavity limits \cite{Megyeri2018a}. The second-order coherence time, determined by Hanbury Brown-Twiss interferometry, is increased over two orders of magnitude as the gain is brought towards threshold. Measurements of cavity ring-down (CRD) show that the photon lifetime, related to the first-order coherence time, is also extended. Finally, prospects for observing the effects of slow light are described.

The experiment has been described in detail elsewhere \cite{Culver,Megyeri2018a,Harvie}, so only a brief outline will be given here. The mode volume of a triangular ring cavity with a finesse of around $1700$ intersects a magneto-optical trap (MOT) of $^{39}$K atoms. The primary cooling transition is between the $|4^2S_{1/2}, F=2\rangle$ ground level and $|4^2P_{3/2},F'=3'\rangle$ ($F$ is the total angular momentum quantum number and primes denote the excited manifold). The relatively small excited-state hyperfine splittings lead to rapid optical pumping out of the cooling cycle and into the $|F=1\rangle$ ground level. Despite the presence of repumping light near-resonant with $|1\rangle\to |F'\rangle$, there remains an effective population inversion relative to $|F=2\rangle$ which leads to Raman gain on $|F'\rangle\to |2\rangle$. In this way, steady-state gain occurs during magneto-optical trapping without the need for additional fields. The amplitude of the gain can be adjusted by changing the flux of the atomic beam feeding the MOT and the intensities and detunings of the cooling and recycling light.

\begin{figure}[h]
\centering
\includegraphics{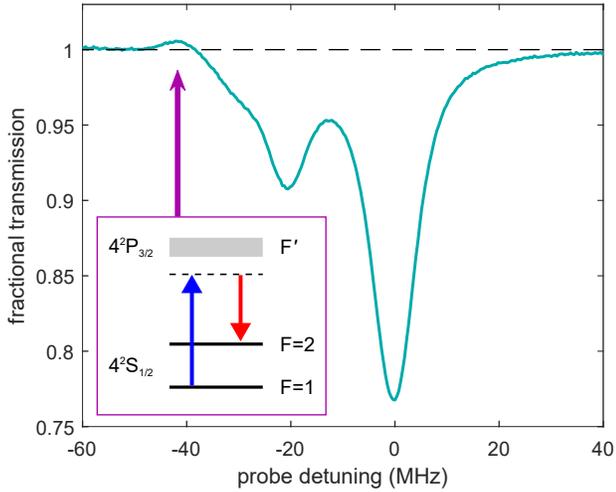}
\caption{(a) Fractional transmission spectrum of the $^{39}$K MOT. Transmission greater than 1 corresponds to gain, and less than 1 to absorption. The probe detuning is referenced to the large $|2\rangle\to |3'\rangle$ absorption resonance; also visible are the barely resolved $|2\rangle\to |2'\rangle$ and $|2\rangle\to |1'\rangle$ features, and the Raman gain peak (indicated by the magenta arrow). The inset shows a simplified energy level scheme, where the MOT repump laser (up arrow) drives gain at the cavity resonance frequency (down arrow) through Stokes scattering.}
\label{fig:gain}
\end{figure}

An example gain spectrum is shown in Fig.~\ref{fig:gain} \cite{data}. The spectrum was measured with the MOT running, using an independent probe laser propagating at a small angle to the cavity axis. This spectrum is intended primarily for illustration --- the peak gain here far exceeds the $0.36\%$ needed to reach the lasing threshold in our system, but as the cavity mode only intersects $\sim 0.5\%$ of the atoms in the MOT, the effects of saturation are not apparent when probing the bulk cloud. Also note the observed linewidth is dominated by relative frequency fluctuations between the probe and Raman pump lasers. The intrinsic linewidth includes homogeneous broadening due to optical pumping and inhomogeneous broadening due to Zeeman and Doppler shifts. These effects vary with the MOT conditions in nontrivial ways, but the gain linewidth $\gamma$ is expected to be of the same order of magnitude as the cold cavity linewidth $\kappa$ in our experiment. Further details of the gain mechanism and transmission measurements were reported in \cite{Harvie}.

The second-order optical coherence of the Raman ring laser was previously measured above and below threshold \cite{Megyeri2018a}. Below threshold, the fluorescence exhibited photon bunching with a coherence time exceeding both of the `natural' decoherence times in the system, namely the radiative lifetime of the upper atomic state (26~ns), and the photon lifetime of the cold cavity ($\kappa^{-1}=86~$ns). We attributed this to the presence of gain, which mitigates the effect of losses at the cavity mirrors. This conclusion is supported by a toy model where the gain medium is treated as a reservoir of inverted oscillators, and is consistent with a more detailed treatment aimed at describing conventional lasers, where the atomic coherence time is explicitly considered negligible compared to the cold cavity lifetime \cite{Haken}. 

To obtain an expression for the coherence time of a below-threshold laser which remains valid as one enters the bad-cavity regime, we first assume the gain medium comprises a large number of statistically independent emitters. Then the first- and second-order coherences are related by $g^{(2)}(\tau)=1+|g^{(1)}(\tau)|^2$, where $\tau$ is the correlation delay time \cite{Loudon}. This implies the first- and second-order coherence times are trivially related. In fact, a calculation following the one in \cite{Laupretre2012} shows that the cavity ring-down signal is proportional to $|g^{(1)}|^2$ after an initial transient delay. In the experiments described here, CRD and HBT therefore effectively measure the same coherence time, $\tau_c$, which can be calculated by considering the photon decay time in the loaded cavity. Ignoring saturation, the small change in mean cavity photon number $n$ following a single pass through the gain medium and around the ring cavity is \cite{QO},
\begin{eqnarray*}
\delta n &=& \mathcal{G}(n+1)-\mathcal{L}n\quad,
\end{eqnarray*}
where $\mathcal{G}$ is the fractional single-pass gain and $\mathcal{L}$ the total cavity losses (including transmission). The time it takes for light to traverse a round trip is $\delta t=\bar{n}_gt_0$ \cite{vanExter,Laupretre2011,Goorskey,Yang,Laupretre2012}, where $\bar{n}_g$ is the group index averaged over the full cavity length and $t_0$ is the cold-cavity transit time. For a simple Lorentzian gain feature with full-width at half-maximum $\gamma$, the arguments presented in \cite{Megyeri2018b} can be adapted to find $\bar{n}_g=1+p\,\kappa/\gamma$, where $p=\mathcal{G}/\mathcal{L}<1$ is the pumping parameter. Approaching threshold ($p\to 1$), this agrees with expressions derived previously for slow-light \cite{vanExter} and superradiant \cite{Bohnet} lasers. Taking $\dot{n}\simeq\delta n/\delta t$, and using $\kappa=\mathcal{L}/t_0$, we finally obtain the steady-state photon number $n_0=p/(1-p)$ and the photon number decay time,
\begin{eqnarray}
\tau_c &=& \frac{1}{\kappa}\,\frac{1+p\,\kappa/\gamma}{1-p}\quad.
\label{eq:tauc}
\end{eqnarray}
It is now clear how gain amplitude and linewidth affect the coherence time. Standard calculations of $\tau_c$ in the context of either laser theory or cavity ringdown assume, explicitly or implicitly, the good-cavity limit ($\kappa\ll\gamma$), for which $\bar{n}_g\to 1$ independent of gain. In the bad-cavity limit, the coherence time is determined solely by the gain amplitude and linewidth.

\begin{figure}[h!]
\centering
\includegraphics{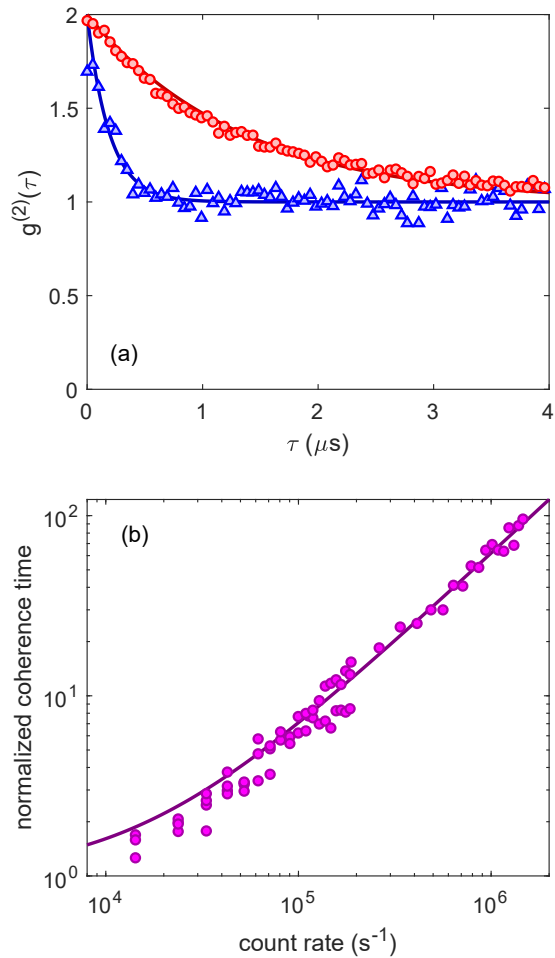}
\caption{Photon bunching in the presence of gain. (a) Second-order coherence for low and high gain (blue triangles and red circles, respectively). The fitted coherence times are $2.18(15)\,\kappa^{-1}$ and $15.1(5)\,\kappa^{-1}$. (b) Normalized coherence time ($\kappa\tau_c$) as a function of mean count rate ($\dot{m}$). The solid curve is a fit to Eq.(\ref{eq:kappatau}), with the slope as the only free parameter.}
\label{fig:HBT}
\end{figure}

To measure $g^{(2)}(\tau)$, the MOT was operated in steady state with the cooling beam intensity reduced to bring the Raman gain below the threshold for lasing. Under these conditions, amplified spontaneous emission circulates in both directions around the ring cavity; there can be several transverse electromagnetic modes but only one linear polarization is resonant for a given cavity length \cite{Culver}. Light propagating in one direction and emitted from one of two identical output mirrors was coupled into a single mode fiber that selectively filtered out all but the TEM$_{00}$ cavity mode. The fiber output was directed to an HBT interferometer consisting of a $50/50$ beam splitter and a pair of single-photon counting module (SPCMs). Photon counts were collected for approximately $60$~s, and lists of timestamps from each counter were recorded. For measurements with very low count rates, photons were collected for longer times of up to $120$~s in order to ensure that enough counts were collected to get reasonable statistics. The lists of time stamps were then split into segments of $5~\mu$s length and the correlation between the two channels was calculated for each segment. Finally, these correlations were averaged over all segments and normalized to the mean count rates to give $g^{(2)}(\tau)$.

The time scale associated with the classical intensity fluctuations of our below-threshold laser ($\sim 1$~ms) is much longer than the typical coherence time, so $g^{(2)}(\tau)$ would not be expected to go from exactly 2 to 1 over short times. A thermal state with slowly varying intensity exhibits a second-order coherence of the form $g^{(2)}(\tau) = A[1+\exp(-\tau/\tau_c)]$, where $A$ depends on the statistics of the long-time fluctuations. This function was fitted to the correlation data with $\tau_c$ and $A$ as free parameters. Figure \ref{fig:HBT}(a) shows examples for low and high gains (plotted in blue and red, respectively). For these sets of data $A\simeq 1$, and the coherence times were $2.18(15)\,\kappa^{-1}$ and $15.1(5)\,\kappa^{-1}$ (numbers in parentheses reflect uncertainties from fits and the previous determination of $\kappa$ \cite{Culver}).

Equation (\ref{eq:tauc}) shows how $\tau_c$ varies with gain, but the values of $\mathcal{G}$ in our experiment are too small to measure directly from spectra such as the one in Fig.~\ref{fig:gain}. Since the steady-state cavity photon number $n_0$ is a monotonic function of $p$, we can use the count rate as a proxy for the gain. The detected count rate is $\dot{m}=\eta\kappa n_0$, where $\eta$ is the total efficiency (including the fraction of cavity light which is transmitted through one output mirror, coupling into the single-mode fibre, losses, and the quantum efficiency of the SPCMs). Expressing Eq.(\ref{eq:tauc}) in terms of the measured count rate, the normalized coherence time is,
\begin{eqnarray}
\kappa\tau_c &=& 1 + \left(1+\frac{\kappa}{\gamma}\right)\frac{\dot{m}}{\eta\kappa} \quad.
\label{eq:kappatau}
\end{eqnarray}
Figure \ref{fig:HBT}(b) shows $\kappa\tau_c$ as a function of count rate. The solid curve is a fit to Eq.(\ref{eq:kappatau}) with the slope as the sole free parameter. The results are in good agreement with Eq.(\ref{eq:kappatau}) over two orders of magnitude in count rate and coherence time. The fitted slope is about twice what we expect from our estimates of the gain linewidth ($\gamma\sim\kappa$) and detection efficiency (we have measured $\eta=1.6\%$, not including an estimated factor of $\sim 0.35$ to account for the fraction of all cavity losses corresponding to transmission through a single output mirror). However, these measurements were performed with low MOT light intensity to keep the gain below the lasing threshold. This reduces the homogeneous broadening, as well as the MOT temperature and radius \cite{Townsend} --- and therefore the Doppler and Zeeman broadening, respectively --- in which case a threshold group index of 4 is reasonable (that is, $\gamma=2\pi\times 600$~kHz). Although the atomic densities here are too low to quantify this via time-of-flight imaging, the agreement between the data and the fit to Eq.(\ref{eq:kappatau}) under the assumption of constant $\gamma$ suggests that the MOT conditions currently present us from resolving slow-light effects.

\begin{figure}[htbp]
\centering
\includegraphics{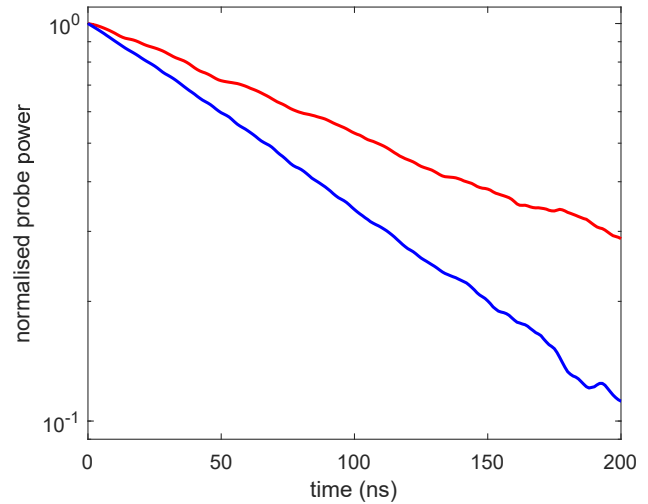}
\caption{Cavity ring-down in the presence of atoms. The upper red curve shows the ring-down curve when the cavity is resonant with the gain peak. An exponential fit (not shown) yields a coherence time of $1.83(7)\,\kappa^{-1}$. The lower blue curve shows a ring-down measurement when the cavity is red-detuned from the gain line by $6.6$~MHz, with $\tau_c=1.04(3)\,\kappa^{-1}$.}
\label{fig:CRD}
\end{figure}

As discussed above, a similar effect should be observed in the first-order coherence time. We measured the photon lifetime --- equal to half the first-order coherence time --- by performing cavity ring-down measurements. While the MOT was running, a relatively bright probe beam (20~nW) was incident on the cavity and the transmitted probe (along with the much weaker fluorescence) was directed to an analog avalanche photodiode (APD). To observe the effect of gain on the decay time constant, the probe was locked on resonance with the gain peak highlighted in Fig.~\ref{fig:gain}. Due to AC Stark shifts induced by the cooling light, the gain peak frequency is not precisely known. Lasing from the intracavity MOT was therefore used as a guide in order to ensure the probe was resonant with the gain. The gain was then reduced until it was back below threshold. Finally, the probe was abruptly switched off using an acousto-optic modulator, and the intensity of the decaying cavity emission was monitored on the APD. This procedure was repeated and averaged 8192 times for each value of gain in order to improve the signal-to-noise ratio. 

Cavity ring-down traces are shown in Fig.~\ref{fig:CRD}. In the presence of gain (red curve) the time constant of the exponential decay, corresponding to $\tau_c$, was $1.83(7)\,\kappa^{-1}$. For comparison, the probe was then red-detuned from the gain resonance by $6.6$~MHz, and the cavity length changed such that transmission was maximized for the new probe frequency. None of the MOT parameters were altered for the second measurement, so that the resonant optical density of the cloud was the same. Figure~\ref{fig:gain} suggests the transmission through the MOT should be near unity at the second probe frequency, and therefore the CRD time constant should approach that of the cold cavity. The result, shown as the blue curve in Fig.~\ref{fig:CRD}, gives $\tau_c=1.04(3)\,\kappa^{-1}$.

To conclude, we have studied the coherence time of a cold-atom laser below threshold. The laser operates in the crossover regime between good- and bad-cavity limits. The second-order coherence was measured in a Hanbury Brown-Twiss interferometer. Photon bunching was observed with a coherence time which varied over two orders of magnitude as the gain approached threshold. The results are in good agreement with a simple analytic formula which shows that the extended coherence under these conditions is dominated by gain rather than dispersion. The first-order coherence was studied through cavity ring-down measurements, which exhibited similar behavior. 

The smallest effects here represent single-pass gains on the order of $10^{-3}$. This suggests that coherence time can be a powerful diagnostic for characterizing extremely small gains. We find HBT superior to CRD in this respect. First, it does not require the addition of a relatively bright probe beam, so that it is essentially non-perturbative. Second, in the absence of the probe, the gain can be estimated via the mean count rate of photons from the amplified spontaneous emission. Finally, the HBT interferometry is more robust with respect to classical intensity noise arising primarily from variations in the MOT, as it is possible to build up histograms from relatively short segments of data. The overall dynamic range of the HBT measurements far exceeded what was possible with CRD in our experiment. 

In the future we would like to observe a greater slow-light effect. The gain linewidth is currently dominated by the conditions of the operating MOT, including intense near-resonant light scattering, residual Doppler broadening, and a distribution of Zeeman states in a spatially-varying magnetic field. By loading the atoms into an intracavity optical dipole trap, it should be possible to reach group indices on the order of $10^4$ \cite{Bohnet}. This would allow us to study the coherence deep into the bad-cavity regime, and to verify theoretical predictions of second-order coherence during steady-state superradiance \cite{Meiser2010}.

\section*{Acknowledgments}
We are grateful to Giovanni Barontini for feedback on the manuscript, and to Vincent Boyer for the loan of equipment and useful discussions.

\section*{Disclosures}
The authors declare no conflicts of interest.



\end{document}